\documentclass[sigconf]{acmart}
\usepackage{booktabs} 
\usepackage{todonotes}
\usepackage{array}
\usepackage{listings}
\usepackage{lstlinebgrd}
\usepackage[shortlabels, inline]{enumitem}
\usepackage{subcaption}
\usepackage{graphicx}
\usepackage{xcolor}
\usepackage{fancyvrb}
\usepackage{verbments}

\usepackage{tabularx}
\usepackage{makecell}
\usepackage{multirow}

\usepackage{xparse}

\usepackage{array}
\usepackage{marginnote}
\let\marginpar\marginnote
\usepackage[noabbrev]{cleveref}

\usepackage{amsmath}
\usepackage{flushend}
\usepackage{ifthen}
\usepackage{standalone}
\usepackage{float}
\usepackage{hyperref}

\setcopyright{none}

\acmConference[MASES'18]{International Workshop on Machine Learning and Software Engineering in Symbiosis}{September 2018}{Montpellier, France}
\acmYear{2018}
\copyrightyear{2018}

\begin{document}
\title{Automatically Assessing Vulnerabilities Discovered by Compositional Analysis}

\author{Saahil Ognawala, Ricardo Nales Amato, Alexander Pretschner, Pooja Kulkarni}
\affiliation{%
  \institution{Technical University of Munich}
  \country{Germany}
}
\email{{saahil.ognawala, ricardo.nales, alexander.pretschner, pooja.kulkarni}@tum.de}

\renewcommand{\shortauthors}{S. Ognawala, R. N. Amato,  A. Pretschner, P. Kulkarni}
\renewcommand{\shorttitle}{Vulnerability assessment with compositional analysis and repository mining}

\begin{abstract}\label{sec:abstract}
Testing is the most widely employed method to find vulnerabilities in real-world software programs. 
Compositional analysis, based on symbolic execution, is an automated testing method to find vulnerabilities in medium- to large-scale programs consisting of many interacting components. However, existing compositional analysis frameworks do not assess the severity of reported vulnerabilities. 
In this paper, we present a framework to analyze vulnerabilities discovered by an existing compositional analysis tool and assign CVSS3 (Common Vulnerability Scoring System v3.0) scores to them, based on various heuristics such as interaction with related components, ease of reachability, complexity of design and likelihood of accepting unsanitized input. By analyzing vulnerabilities reported with CVSS3 scores in the past, we train simple machine learning models. By presenting our interactive framework to developers of popular open-source software and other security experts, we gather feedback on our trained models and further improve the features to increase the accuracy of our predictions.
By providing qualitative (based on community feedback) and quantitative (based on prediction accuracy) evidence from 21 open-source programs, we show that our severity prediction framework can effectively assist developers with assessing vulnerabilities. 


\end{abstract}

%
%

\ccsdesc[500]{Security and privacy~Vulnerability management}
\ccsdesc[500]{Software and its engineering~Software testing and debugging}


\maketitle

\section{Introduction}\label{sec:introduction}
Due to an increase in the size, features, and complexity of software applications coupled with increased automation in the hands of professional bug hunters, we have seen an explosion in the number of vulnerabilities exposed in popular software. Automated software testing is the preferred way for early detection of bugs in programs leading to vulnerabilities. Popular vulnerability scanners, however, report many, so-called, \emph{false positives} \cite{torchiano2010assessing} that may never materialize in a real-world usage of the program or associated components. In addition to expert knowledge, this calls for vulnerability assessment techniques for reported vulnerabilities that take into account the context of development, usage, underlying assets and the likelihood of exploitation \cite{christakis2016developers}. 

One such automated testing and vulnerability discovery tool is Macke \cite{ognawala2016macke}. Macke is a compositional analysis tool based on symbolic execution, that achieves higher instruction coverage and discovers more potential vulnerabilities in many open-source programs than forward (simple and without compositional analysis) symbolic execution tools, such as KLEE \cite{cadar2008klee}. The basic idea behind Macke is symbolic execution of isolated components in a program, summarizing vulnerabilities found in them and, performing a reachability analysis for the discovered vulnerabilities to generate exploits. However, when exploits cannot be generated (potential false-positives), Macke reports vulnerabilities without providing any contextual information to help developers prioritize the fixing process for reported vulnerabilities.

Fortunately, the Common Vulnerability Scoring System (CVSS) \cite{cvss3}, a standard that is being rapidly adopted by the IT industry, is an existing system to rate the severity of a vulnerability and, hence, prioritize them. CVSS scores vulnerabilities by combining some properties of a vulnerability through empirically derived parameters. 

In this paper, we will present a data mining and machine learning based technique for correlating features of vulnerable components discovered by compositional analysis in C programs and predicting CVSS3 base-score values. 

\paragraph{Problem:} 
The state-of-the-art vulnerability scanners do not satisfactorily report found vulnerabilities.
Static analysis tools, such as Splint, report too many false-positives. 
Compositional analysis tools, such as Macke, report fewer vulnerabilities (and, hence, possibly fewer false-positives) than static analysis tools but, in the absence of contextual information, it is difficult to triage bugs and prioritize reported vulnerabilities for which no exploit could be generated. 

\paragraph{Solution:} 
We collect a set of vulnerabilities reported in the past, with CVSS3 scores, and the corresponding versions of the programs affected by them. We compositionally analyze these programs with Macke and process the results to extract some features from them. Then, using scores of the collected set of past vulnerabilities as ground truth, we apply basic machine learning techniques to learn a model that can predict CVSS3 scores for vulnerable components from the features. We present the vulnerabilities with predicted severities to security and software testing experts and gather feedback on our prediction framework. Using this feedback, we add more features and obtain new models that predict severities for reported vulnerabilities with higher accuracy. 

\paragraph{Contribution:} Closing a gap in this field, we present a novel strategy to utilize the contextual information provided by compositional analysis that affects the severity of discovered vulnerable functions in C programs. Building on the empirical evidence that compositional symbolic execution reveals more vulnerabilities than forward symbolic execution \cite{ognawala2016macke}, we state and evaluate the claim that it also results in relevant information for prioritizing vulnerabilities, such as ease of exploitation and the extent of damage that a vulnerability might cause if exploited. To the best of our knowledge, ours is the first framework to automatically assess severity of vulnerabilities reported by compositional symbolic execution, and evaluate it based on the accuracy of assessment and feedback from the open-source community. 

This paper is structured as follows -- \cref{sec:background} gives an overview of the background to our work. In \cref{sec:methodology}, we describe our research methods and implementation of our prediction framework. In \cref{sec:evaluation}, we list and describe the evaluation criteria for our methods, results and their interpretation and threats to validity of our experiments. \Cref{sec:related-work} lists some past works related to ours and in \cref{sec:conclusion} we conclude the paper. 
\section{Background}\label{sec:background}

\subsection{Symbolic Execution}\label{sec:background-symbolic-execution}
Symbolic execution was introduced as a technique for software testing by \citeauthor{king1976symbolic} \cite{king1976symbolic}. It is a deterministic method that uses instrumentation to dynamically collect constraints representing branching conditions in a program and solves these path constraints to generate inputs that execute the corresponding paths in the program. A path constraint (or \emph{path condition}) may be defined as an ordered sequence of conditional branches that a program's execution takes to reach from an entry point (e.g.\ \texttt{main} function) to an exit point (e.g.\ \texttt{return} statement) of the program. Some of these paths may lead to unhandled exceptional behaviour in the program (such as buffer-overflows). Symbolic execution, and its practical approaches such as \emph{concolic execution} \cite{sen2005cute} and \emph{whitebox fuzzing} \cite{godefroid2008automated}, have been shown \cite{godefroid2008automated} to be capable of extracting complex path-conditions for edge-cases and exceptions where other methods, like random testing, have failed to find potential vulnerabilities. 

However, symbolic execution and its variants suffer from bottlenecks of underlying constraint solvers \cite{erete2011optimizing} and path-explosion \cite{cadar2013symbolic}, which leads to a major degradation of its performance in terms of both, path-coverage and vulnerability discovery. 

\subsection{Compositional Analysis with Macke}\label{sec:background-macke}
Compositional analysis or compositional symbolic execution has been proposed \cite{christakis2015ic,ognawala2016macke,pretschner2003compositional} as a mitigation strategy for the path-explosion problem in simple symbolic execution. The basic idea of compositional analysis is as follows -- instead of symbolically executing the full program, we symbolically execute all components, such as functions, that can be executed in isolation. 
Then, we only focus on the inter-compositional interactions of these components, by means of directed symbolic execution \cite{ma2011directed}. In this way, the symbolic execution engine would not have to deal with all those paths that do not constitute any communication-related instructions between the components. 
Macke \cite{ognawala2016macke} is such a compositional analysis tool for C language programs, where the components are functions. 
We will, briefly, formally describe the working of Macke. 

All the following description applies to all function, $f$, in the program, $P$, that are \emph{executable} and we assume that there is at least one such function in $P$. An executable function is defined as a function that can be called \emph{with arguments} from another function in the program. Let, the set of all executable functions in $P$ be $F_P$. 
An executable function, $f$, may be represented as an \emph{execution tree} that characterizes the possible paths followed during a symbolic execution of the function, as described by \citeauthor{king1976symbolic} \cite{king1976symbolic}. If the list of arguments to the function, $f$, is $I = \{\alpha_1, \alpha_2, \dots \alpha_n\}$, then a path-condition, $pc$, is a Boolean expression over $\alpha_i$'s. For every symbolic execution of a function
\begin{align}\label{eq:pc-initial}
    pc_{initial} = True
\end{align}
Whenever a branching statement, e.g.\ \texttt{If-else}, with branching condition, $q$, is encountered $pc$ is extended as follows
\begin{align}\label{eq:pc-addition}
    pc = pc \land r
\end{align}
where, $r = q$ or $\lnot q$. 

Let the set of all $pc$'s symbolically executed in an isolated function, $f$, be $PC_f$ \footnote{Note that $PC_f$ does not contain a set of all \emph{possible} paths in $f$, but only those that were symbolically executed.}. By the end of every execution, a $pc$ will represent a path in $f$ that ends due to
\begin{enumerate*}
    \item \texttt{return} statement, or
    \item program crash, or
    \item exceptional condition, such as an assertion failure.  
\end{enumerate*}
Macke considers case (2) as a segmentation fault occurring due to \emph{buffer-overflow} vulnerability and case (3) as a special case of \emph{assertion-failure} vulnerability. In this paper, we will only focus on buffer-overflows. Let the subset of $PC$ that corresponds to all buffer-overflow vulnerabilities in $f$ be $PC^{vuln}$.  The corresponding arguments that execute the paths in $PC^{vuln}$ can be considered exploits for these vulnerabilities. 

After symbolically executing all functions, $f$, in isolation Macke performs directed symbolic execution to confirm the reachability of vulnerable paths via function calls. 

For describing directed symbolic execution and reachability let us first define a function, $parents$. 
For any two isolated functions, $f_1, f_2 \in F_P$
\begin{align}
    \forall f_1, f_2 \in F_P: f_1 \in parents(f_2) \iff f_1 \mathbf{calls} f_2
\end{align}

I.e.\ The $parents(f)$ function lists all functions that may \emph{potentially} call $f$. 

Now, let $f_1 \in parents(f_2)$ and the set of vulnerabilities discovered in $f_2$ be $PC^{vuln}_{f_2}$. Then, a vulnerability in $f_2$ is said to have \emph{infected} function, $f_1$ if
\begin{align}\label{eq:compositional-implication}
    \exists pc_{f_1} \in PC_{f_1}, \exists pc_{f_2} \in PC^{vuln}_{f_2}: pc_{f_1} \land pc_{f_2}
\end{align}
Using a directed search strategy \cite{ma2011directed,majumdar2009reducing}, Macke reduces the set of paths by removing those paths in $PC_{f_1}$ that do not satisfy the relation in \cref{eq:compositional-implication}, thereby reducing path-explosion. 
Additionally, it reports those vulnerabilities in isolated functions that cannot be found by forward symbolic execution of the $main$ function but, nonetheless, \emph{might} be reproducible through some functions in $parents(f)$. 
The above properties of Macke will be used for machine learning, as described in \cref{sec:methodology}. 

\subsection{CVSS3}\label{sec:background-cvss3}
The first version of Common Vulnerability Scoring System (CVSS) \cite{mell2007complete}, introduced in 2003, proposed a way to capture the principal characteristics of a vulnerability and assign a numerical score reflecting its severity. The latest version to be used is version 3.0, or simply CVSS3. CVSS3 \cite{cvss3} consists of a base-score and optional temporal and environmental scores. In this paper, we will only focus on predicting the base-score. CVSS3 base-score is made up of the following metrics, which can take one of the corresponding values
\begin{itemize}
    \item\emph{Attack Vector (\texttt{AV})} -- The context in which exploiting the vulnerability is possible. \textbf{Allowed values:} \emph{network}, \emph{adjacent}, \emph{local} and \emph{physical}. 
    \item\emph{Attack Complexity (\texttt{AC})} -- The complexity of the attack process, if possible. \textbf{Allowed values:} \emph{low} and \emph{high}. 
    \item\emph{Privileges Required (\texttt{PR})} -- The level of privileges an attacker must have to carry out an exploit. \textbf{Allowed values:} \emph{none}, \emph{low} and \emph{high}. 
    \item\emph{User Interaction (\texttt{UI})} -- The amount of direct user interaction required for the attacker to carry out an exploit. \textbf{Allowed values:} \emph{none} and \emph{required}.  
    \item\emph{Scope (\texttt{S})} -- Whether or not other components (changed scope) than the vulnerable one can be affected if the vulnerability is exploited. \textbf{Allowed values:} \emph{unchanged} and \emph{changed}. 
    \item\emph{Confidentiality (\texttt{C})} -- The amount of confidential data that will be exposed if the vulnerability is exploited. \textbf{Allowed values:} \emph{none}, \emph{low} and \emph{high}. 
    \item\emph{Integrity (\texttt{I})} -- How much information can the attacker modify in the exploited component. \textbf{Allowed values:} \emph{none}, \emph{low} and \emph{high}. 
    \item\emph{Availability (\texttt{A})} -- Can the attacker deny access to the exploited component and whether that component is critical. \textbf{Allowed values:} \emph{none}, \emph{low} and \emph{high}. 
\end{itemize}
In \cref{sec:evaluation}, we will see how effectively we can predict all of the above CVSS3 base-score metrics. 

\subsubsection*{Comparison with Bugzilla severity}
The popular bug-reporting platforms such as \cite{gnome-bugzilla} use \emph{Bugzilla}'s nominal categories for prioritizing bug fixes. However, in past works \cite{ayari2007threats} and from our own experience, we note that the context-free ranking system of Bugzilla results in an order that doesn't truly represent the severities of bugs. The first reason for this is that, for most of the bugs analyzed by us (\cref{sec:evaluation}), the development community ignored the ``priority'' field of the reports, and used only the ``severity'' field as a proxy for both, priority and severity. Unlike Bugzilla, for calculating CVSS and CVSS3 scores, evaluation \emph{all} base-score values is mandatory. Secondly, we also found many instances in the analyzed programs where the severity values in Bugzilla were changed by the developers when, either, the underlying assets were not considered important enough, or the bug would not be fixed because it was too complex to exploit it. However, CVSS base-scores explicitly take into account underlying assets and attack complexity, thereby eliminating the need for further arbitrary adjustment. 
\section{Methodology}\label{sec:methodology}
In this section, we will give a high-level overview of the implemented methods, including data collection, programs' compositional analysis, initial feature extraction, the first stage of machine learning, interactive feedback gathering and machine learning with combined features.

In the following subsections, we will use the running example of \emph{Autotrace 0.31.1}, a UNIX command-line-based program to convert bitmap image formats to a vector graphics format. 

\subsection{Data Collection}\label{sec:data-collection}
We, firstly, require a set of known vulnerabilities with CVSS3 base-scores to learn a predictive model. For this study, we only consider \emph{buffer-overflow vulnerabilities}. 
Because of reasons listed in \cref{sec:background-cvss3}, Bugzilla repositories, such as \cite{gnome-bugzilla}, were not suitable for our study. 
The National Vulnerability Database (NVD) \cite{nvd} contains analyses of Common Vulnerabilities and Exposures (CVEs) by aggregating program and vulnerability description, vulnerability type enumeration, applicability statements, impact metrics (in CVSS or CVSS3 format) and other relevant references. 

After selecting NVD as our preferred database, we needed to filter the data for reports
\begin{enumerate}
    \item that follow CVSS3 notation, instead of CVSS 1.0, 
    \item about C programs, 
    \item about buffer-overflows, and
    \item that state the name of the vulnerable function.
\end{enumerate}

Setting the filters on the data collected from NVD, as described above, we were left with a set of vulnerability reports that were thorough in their description and had CVSS3 scores attached to them. The next step is to, manually or automatically, determine the name and version of the affected program and download the source-code of it. In \cref{sec:eval-vulnerability-reports}, we will present the size of dataset and ground-truth for machine learning. 

\subsection{Compositional Symbolic Execution}\label{sec:compositional-symbolic-execution}
As described in \cref{sec:background-macke}, Macke performs compositional symbolic execution by symbolically executing isolated functions in a program and, then, performing a reachability analysis for the reported vulnerabilities from parent (calling) functions. In this step of our methodology, we perform compositional analysis on the programs collected in \cref{sec:data-collection}, using Macke. 

\begin{figure*}
    \includegraphics[width=\linewidth,height=0.25\linewidth]{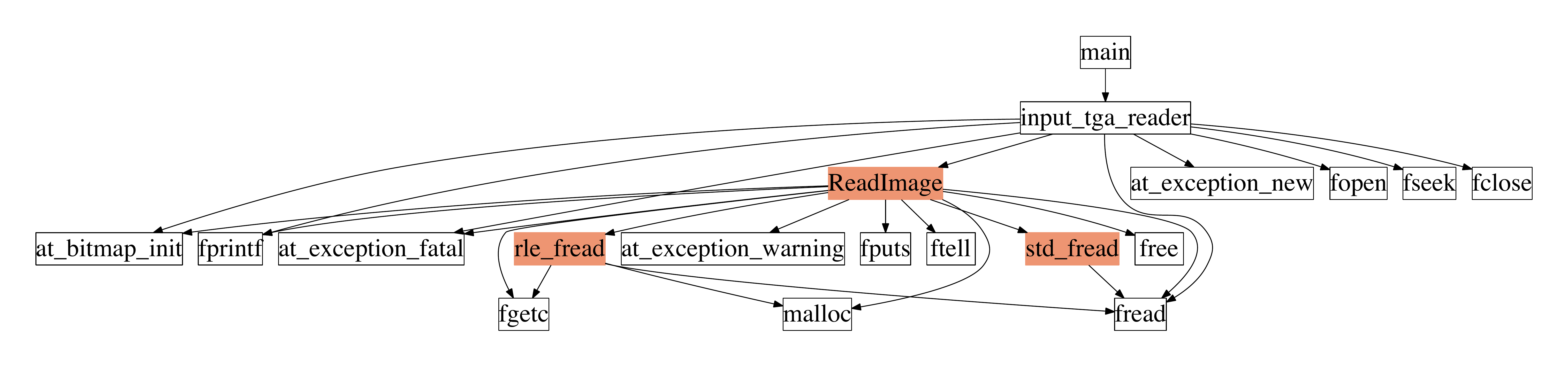}
    \caption{Call-graph of Autotrace 0.31.1 program to convert a TGA bitmap to vector graphics format}
    \label{fig:input-tga-autotrace}
\end{figure*}

The output of Macke includes a JSON file that lists
\begin{enumerate}
    \item all discovered vulnerabilities, 
    \item source-code location of the vulnerable instruction, and
    \item functions through which an identical vulnerability may be exploited. 
\end{enumerate}
The results of the analysis with Macke are presented in \cref{sec:eval-macke-analysis}. In addition to the above list, Macke also outputs a \emph{call-graph} of the program. The call-graph for Autotrace 0.31.1 is shown in \cref{fig:input-tga-autotrace}. The vulnerable functions, as discovered by Macke, in this graph are highlighted. 

%
%

\subsection{Feature Extraction}\label{sec:feature-extraction}
The output of running Macke on the candidate programs are processed in this step to extract some features related to the vulnerabilities and vulnerable functions. We will now describe these features and our intuition behind including them as possibly correlating factors for predicting the severity of vulnerabilities. Please note that in the following list the terms ``nodes'' and ``functions'' are used interchangeably. Also, we assume that each function in the call-graph has an equal likelihood of sanitizing an input (argument). 

\begin{enumerate}
    \item \emph{Node degree (\texttt{d\_in}, \texttt{d\_out})}, defined as the number of callers or callees (called, henceworth, also as \emph{neighbours}) for a function in the call-graph. As explained by \citeauthor{el2001prediction} \cite{el2001prediction}, and \citeauthor{nagappan2006mining} \cite{nagappan2006mining}, node degree is an important feature because the higher the node's degree, the more likely that a vulnerability in it may infect other functions, thereby leading to a failure in the program.
    For the function \texttt{rle\_fread} (\cref{fig:input-tga-autotrace}) the values of incoming node degree, \texttt{d\_in}, and outgoing node degree, \texttt{d\_out}, are $1$ and $3$, respectively. For function, \texttt{std\_fread}, \texttt{d\_in} and \texttt{d\_out} are $1$ and $1$, respectively. 
    
    \item \emph{Distance to interface (\texttt{di})}, defined as the length of the shortest path from an interface (such as \texttt{main} function) to the function. The shorter the distance from an interface, the less likely it is that a pointer argument was sanitized before being accessed. 
    The value of \texttt{di} for \texttt{std\_fread}, as seen from \cref{fig:input-tga-autotrace} is $3$, while it is $2$ for \texttt{ReadImage}. 
  
    \item \emph{Clustering coefficient (\texttt{cc})}, defined as the ratio of neighbouring functions of a node that are also mutually connected (as caller-callee pair).
    The intuition behind this feature is that the bigger a \emph{cluster} in which a vulnerable function is, the more likely it is that this vulnerability may be exploited by another function in the cluster. 
    From \cref{fig:input-tga-autotrace}, we can see that \texttt{rle\_fread} has 4 immediate neighbouring functions (callers or callees). Out of $6$ possible pairs of neighbours of \texttt{rle\_fread}, $3$ are also connected to each other in a caller-callee relationship. Therefore, the value of \texttt{cc} for \texttt{rle\_fread} is $0.5$.
    
    \item \emph{Node path length (\texttt{nl})}, defined as the average number of steps that need to be taken to reach any reachable node from the function being analyzed. If the node path length for a node is low, it denotes a higher likelihood that an argument leading to buffer-overflow is \emph{not} being sanitized before being passed between functions, assuming every function sanitizes input with the same likelihood. 
    From \cref{fig:input-tga-autotrace}, we can see that there are three functions that are reachable from \texttt{rle\_fread} and their distances from \texttt{rle\_fread} are $1$, $1$ and $1$, respectively. Hence, the value of \texttt{nl} for \texttt{rle\_fread} is $1$ (average of $1$, $1$ and $1$). In our implementation, we detect loops in the call-graph and stop counting nodes when we encounter a loop (e.g.\ in recursion). 
    
    \item \emph{Vulnerabilities discovered (\texttt{nv})} depends on the output of Macke, and it denotes the number of unique vulnerable instructions discovered by Macke. A higher number of buffer-overflow vulnerabilities indicate \cite{nagappan2005static} that an argument might be assumed to be sanitized by a calling function, but such a sanitization never took place in reality. 
    In the functions, \texttt{rle\_fread} and \texttt{std\_fread}, Macke discovered $1$ vulnerable instruction each, i.e.\ \texttt{nv} is $1$. However, for \texttt{ReadImage} (\cref{fig:input-tga-autotrace}) Macke found the instructions in, both, \texttt{rle\_fread} and \texttt{std\_fread}, to be exploitable. Therefore, \texttt{nv} is 2 for \texttt{ReadImage}. 
    
    \item \emph{Maximum length of infection (\texttt{li})} depends on the results of compositional analysis. 
    It is the longest chain of \emph{caller-callee} pair through which the same vulnerability was discovered by Macke. The intuition behind this feature \cite{ognawala2016macke} is that if the same vulnerability can be exploited in a long chain, then it is more likely that a sanitization on passed arguments was not performed.
    We can see from the call-graph in \cref{fig:input-tga-autotrace} that the underlying ``causes'', or unsafe operation, for the vulnerabilities in \texttt{ReadImage} function are in \texttt{std\_fread} and \texttt{rle\_fread}. In both cases, the maximum length of infection, \texttt{li}, is $2$ ($ReadImage \rightarrow std\_fread$ or $ReadImage\rightarrow rle\_fread$). 
\end{enumerate}

\subsection{Prediction of CVSS3 Base Scores}\label{sec:predicting-cvss3}
\subsubsection{Preparation of data for prediction models}
With the ground truth and features related to functions and vulnerabilities, the next step is to train machine learning models to learn the correlation between these features and CVSS3 base-scores and use a linear or non-linear combination of the features to predict these base-scores. 

We consider each CVSS3 base-score value as individual targets for prediction and generate the final CVSS3 severity score, $severity$, based on the formula in the original specification document of CVSS3 \cite{cvss3}, $calc\_cvss3$. i.e.\vspace{-1em}
\begin{align}\label{eq:cvss3-base-score}
    severity = calc\_cvss3(y_{AV}, y_{AC}, y_{PR}, y_{UI}, y_{S}, y_{C}, y_{I}, y_{A})
\end{align}
where the subscripts, $base$,  are the CVSS3 base-scores described in \cref{sec:background-cvss3}, 
Concretely, \vspace{-1em}
\begin{align}\label{eq:y-base}
    y_{base} = f^{L}_{base}(V)
\end{align}
where,\vspace{-1em}
\begin{align}\label{eq:X-array}
    V = [d\_in, d\_out, di, cc, nl, nv, li]
\end{align}

The superscript, $L$, in \cref{eq:y-base} stands for ``learned'', denoting the learned model, $f^{L}_{base}$. The $[\dots]$ in \cref{eq:X-array} represents a vector of feature values.  

\subsubsection{Machine learning models}
We apply two standard machine learning algorithms to learn functions, $f^{L}$, viz.
\begin{enumerate}
    \item Random-forest classifier,
    \item Naive Bayes classifier, 
\end{enumerate}
We use scikit-learn \cite{scikit-learn}, a widely used machine learning toolkit in Python. For all models, we apply K-fold cross-validation on the training dataset. We, then, apply the model with best validation score on the test dataset to calculate the test scores, which are reported in \cref{sec:evaluation}. 

For presentation to the experts for gathering feedback, we apply the best of all machine learning models (based on test scores) to predict scores for those vulnerabilities that were discovered by Macke but were previously unreported. 

\subsection{Presentation and Feedback Gathering}\label{sec:presentation-feedback-gathering}
\subsubsection{Interactive reporting of vulnerabilities}\label{sec:interactive-reporting}
\begin{figure}
    \includegraphics[width=1.1\linewidth,keepaspectratio=true]{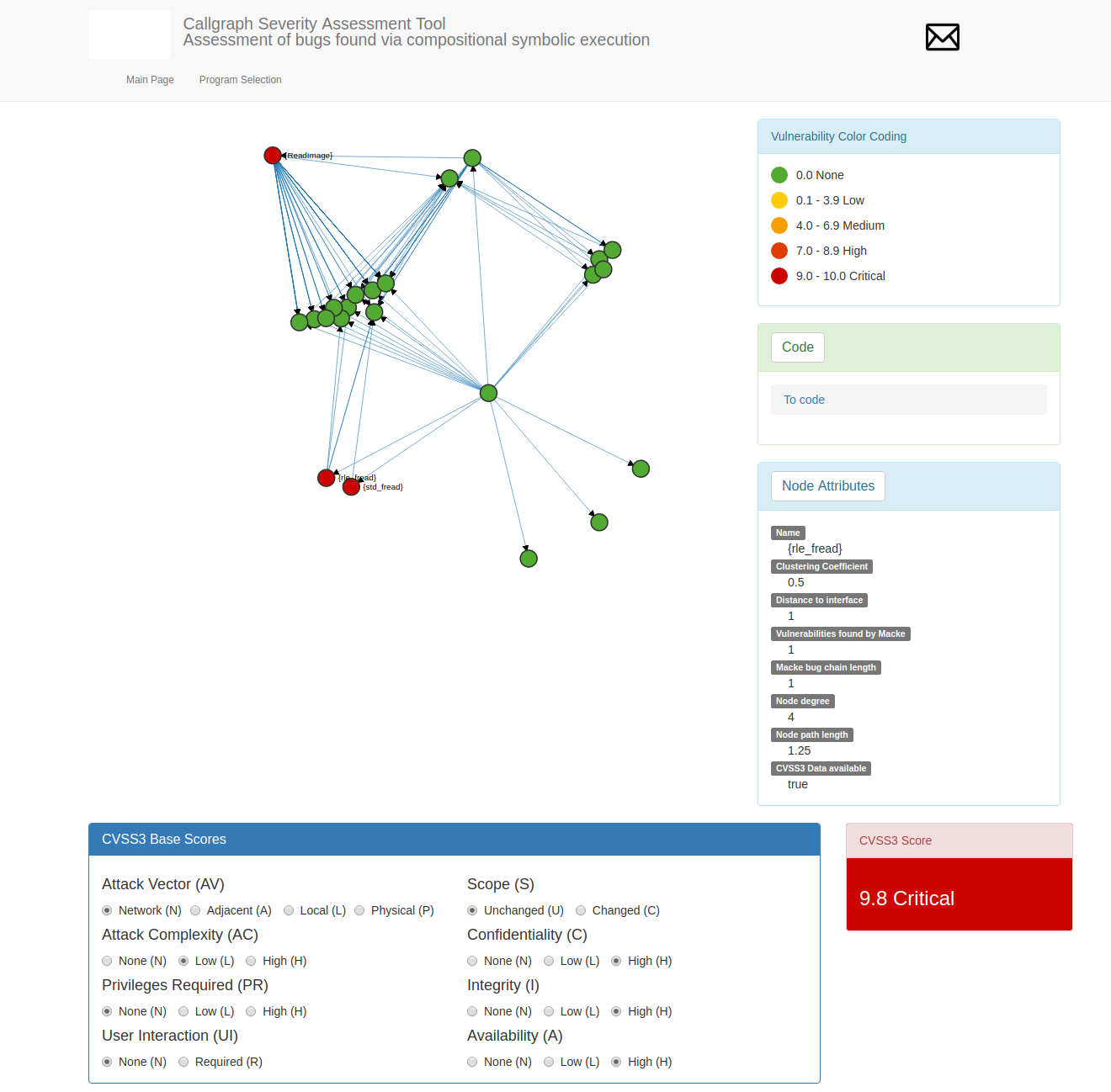}
    \caption{Severity assessment interface for Autotrace 0.31.1 with interactive call-graph\vspace{-2em}}
    \label{fig:callgraph-and-cvss}
\end{figure}

After learning predictive models for CVSS3 base-score values we present the predictions to security and software development experts in an interactive medium to obtain feedback from them. 
The requirements for such a presentation medium are
\begin{enumerate}
    \item Ability to interact with the call-graph, including zooming in on functions and viewing the source-code. 
    \item Ability to view CVSS3 base-scores and aggregate values, by clicking on the function node. 
    \item Ability for the user to change base-score values, using a numerical text-box. The aggregate CVSS3 value must be updated automatically.  
    \item Tracking all the above interaction, including when a function node is clicked, source-code is expanded and a base-score value is updated. 
    \item Ability for the user to send textual feedback, using a multi-line text-box. 
\end{enumerate}
We created a web-application on a server running NodeJS, with a ReactJS frontend. The resultant interface, satisfying all the requirements, for Autotrace program is shown in \cref{fig:callgraph-and-cvss}. 

\subsubsection{Feedback from experts}
The goal of this step is to learn from our target audience how effective a severity prediction tool would be in improving the manual task of prioritizing vulnerabilities. 
We provide a link to the online tool to selected experts listed in \cref{sec:eval-feedback} and show the predicted CVSS3 base-score values and the aggregate score, for previously reported and unreported vulnerabilities. However, the distinction between previously reported and unreported vulnerabilities is invisible to the experts, so that their feedback may be validated. For previously unreported vulnerabilities, if the experts disagree with the predicted values, we ask for their reasons for disagreement, so that we can improve our prediction framework. 
From the experts that choose to participate in our survey, we collect the following information
\begin{enumerate}
    \item Which function node was expanded (by clicking on it, as shown in \cref{fig:callgraph-and-cvss}).
    \item Function nodes for which any CVSS3 base-score value was changed (if an expert disagreed with the predicted value), along with the old and new values, 
    \item Function for which the expert referred to the source-code, 
    \item Textual feedback, when the expert wanted to provide feedback or clarifications, 
    \item (optional) expert's full name and email if they agreed to be contacted by us. 
\end{enumerate}
Above information is stored in a MySQL backend and processed manually by us at the end. 
The qualitative analysis of received feedback will constitute our first measure of effectiveness. 

\subsection{Feature Addition}\label{sec:feature-addition}
Based on the feedback received from experts, we add more features to extend the predictive power of our machine learning models. The new features added and the intuition behind including them are as follows

\begin{enumerate}
    \item \emph{Function size (\texttt{s})}, a simple count of LLVM \cite{lattner2004llvm} instructions in the function. 
    The intuition behind this feature is that a higher number of instructions might indicate that, instead of splitting the functionality across various single-purpose function, it exists in a singular function. Vulnerability inside such a function that is not interacting with, and hence doesn't depend for the sanitization of its inputs on, other functions of the program should be fixed with a high priority. 
    
    \item \emph{Approximate function complexity (\texttt{fx})}, a count of LLVM \emph{basic-blocks} in the function. A basic-block \cite{hennessy2011computer} is defined as a straight-line of instruction-sequence that contains no branches inside it, other than the entry or exit points of the block. 
    \citeauthor{nagappan2006mining} \cite{nagappan2006mining} claimed and showed with case studies that the number of basic blocks and arcs in CFG, which may be an approximation of McCabe's complexity measure, correlate well with the possibility of a module's failure in many projects. In our case, a higher function complexity (in terms of basic-blocks) may indicate a higher likelihood that a vulnerable instruction was left unhandled unintentionally. 
%
    
    \item \emph{Pointer parameters (\texttt{pt})}, the number of parameters that a function accepts that are of pointer type. Pointers are important for a vulnerable function because we are dealing with buffer-overflow vulnerabilities only. If more pointer parameters are specified, then there is a higher likelihood that access to at least one of them might be in an unsafe manner. 
    \end{enumerate}
After adding these features to the existing features, we learn the new predictive functions, $f^{L}$, for all CVSS3 base-score values, using the same machine learning algorithms as listed in \cref{sec:predicting-cvss3}. The accuracy of the final learned model, on previously reported vulnerabilities, constitutes the second effectiveness measure of our tool.  

\section{Evaluation}\label{sec:evaluation}
Based on the methodology of our study, as described in the previous section, we will now discuss the experiments carried out for evaluating the effectiveness of our prediction framework. The raw-data related to all stages of evaluation has been made public by us at \cite{osf2018cvss3}. 

\subsection{Vulnerability Reports and Affected Programs}\label{sec:eval-vulnerability-reports}
The \emph{21 open-source programs} analyzed in this study are listed in \cref{tab:programs-analyzed}. The second column in \cref{tab:programs-analyzed} lists the lines of source-code in the programs. The analyzed programs range from, approximately, 2,000 to 470,000 lines of code, with an average of, approximately, 100,000 lines of code

The third column in \cref{tab:programs-analyzed} lists the number of \emph{connected} functions in the analyzed programs. 
We define connected functions as the functions that are (potentially) reachable, as determined by a simple syntactic analysis, from an entry point in the program. 
Many of the analyzed programs have several direct entry points (such as \texttt{main}) or public APIs. In this study, for every program, we only choose the connected component that contains a function which is present in one of the collected vulnerability reports (from NVD). 

The fourth column in \cref{tab:programs-analyzed} lists the number of vulnerable functions (functions containing at least one reported vulnerability) reported in the respective program in NVD. These vulnerability reports are obtained after applying all the filters described in \cref{sec:data-collection}. The number of CVSS3 scored vulnerable functions, as we can see from \cref{tab:programs-analyzed}, is not large enough to effectively train any machine learning algorithm. Therefore, to augment this list obtained from NVD, we manually score some more vulnerable functions, listed in the fifth column of \cref{tab:programs-analyzed}, that are discovered by Macke.
\begin{table}[]
    \centering
    \caption{Programs analyzed and vulnerabilities in them}
    \label{tab:programs-analyzed}
    \begin{tabularx}{1.1\linewidth}{@{}|>{\raggedright}X|r|r|r|r|r|@{}}
        \hline
        \thead{Program \\and Version} & \thead{LOC} & \thead{Connec-\\ted \\ func-\\tions} & \thead{Vulner-\\able \\ func-\\tions \\ from\\ NVD, \\(with \\ CVSS)\\ $||N||$} & \thead{Vulne-\\rable \\ func-\\tions \\ CVSS \\ scored \\ manually,\\ $||M||$} & \thead{Vulner-\\able \\ func-\\tions \\ found \\ by \\Macke, \\ $||X||$} \\ \hline
        BlueZ 5.42                             &    286,206 &        49            &     3    & 2                 & 6                                                \\
        AutoTrace 0.31.1                       &  18,581   &       23             &  3     & 0                  & 3                                                \\
        GraphicsMagick 1.3                   &   324,422  &     22               & 4 & 4                          & 10                                                \\
        Icoutils 0.31.1                        &   40,093  &               45     & 2    & 3                     &  5                                               \\
        ImageMagick 6.0.4-8                    &  476,747   &         51           & 1     & 3                         & 8                                                \\
        Jasper 1.900.27                        &  46,578   &        33            &   3     & 4                     &   19                                              \\
        Jasper 2.0.10                          &  46,622   &          33          &       2     & 3              &  6                                               \\
        Libarchive 3.2.1                      &  204,993   &       62             &  1         & 4               &  15                                               \\
        Libass 0.13.3                          &  18,745   &       46             &  1      & 3                  &  29                                               \\
        Libmad 0.15.1                          &  12,866   &      22              &  1     & 1                   & 4                                                \\
        Libplist 1.12                          & 6,075    &        69            &  1     & 5                   & 27                                                \\
        Libsndfile 1.0.28                      &  85,189   &    153                & 1     & 3                    &  40                                               \\
        Libxml2 2.9.4                          &  334,796   &      36              & 2     & 3                 & 22                                                \\
        Lrzip 0.631                            &  18,622   &       115             &  1      & 1                  & 7                                                \\
        Openslp 2.0.0                          &  55,545   &    27                & 1      & 3                   & 17                                                \\
        Potrace 1.12                           &  12,928   &    28                & 1        & 3                 & 13                                                \\
        Rzip 2.1                               &  2,651   &         34           & 1         & 3                & 19                                                \\
        Tcpdump 4.9.0                          &  103,152   &   13                 & 1    & 1                     & 7                                                \\
        Tiff 4.7.0                            &  82,725   &         125           & 3      & 2                   &  6                                               \\
        Virglrenderer 0.5.0                    &  57,213   &    70                & 2    & 0                     & 21                                                \\
        Ytnef 1.9.2                            &  4,818   &        70            & 6      & 1                   &  12                                              \\
        \hline
        & & \textbf{Total} & \textbf{41} & \textbf{52} & \textbf{296} \\
        \hline 
    \end{tabularx}
\end{table}
\subsection{Analysis with Macke}\label{sec:eval-macke-analysis}
After collecting the vulnerability reports and the respective affected versions of the programs, the next step is to apply Macke on these programs to find the same vulnerabilities, or more, that are listed in \cref{sec:eval-vulnerability-reports}.

The sixth column in \cref{tab:programs-analyzed} lists the number of vulnerable functions discovered by Macke in approximately \emph{30 minutes}. We can see that the number of unique vulnerable functions found by Macke, 296, is more than the reported vulnerable functions in NVD. After carefully analyzing the results, we found that \emph{all vulnerabilities reported in NVD} were also found by Macke within the given time-limit. The extra vulnerabilities and vulnerable functions may or may not be true positives. 

\subsection{Feature Extraction and Machine Learning}\label{sec:eval-feature-extraction-and-machine-learning}
After running Macke on the programs to be analyzed, the next step is to extract features from the programs and their analyses results, as described in \cref{sec:feature-extraction}, and learn our prediction models using them. 
The \emph{ground truth} to be used for machine learning is $G = N \cup M$, where $N$ and $M$ are as shown in \cref{tab:programs-analyzed}. 

As described in \cref{sec:predicting-cvss3}, we trained two machine learning models for learning the CVSS3 base-score values using features listed in \cref{sec:feature-extraction}. The dataset for learning, $G$, is split into training ($75\%$) and testing ($25\%$) sets. 
The training set is, then, split for 4-fold cross-validation, i.e.\ 4 machine-learning models are trained by holding out each of the 4 folds one-by-one, and the best of these 4 models chosen. Because of the small training set, we chose only 4 folds for cross-validation, instead of, say, 10 folds that is more common in many papers. 
To remove the effect of random initial states, for all iterations, we train 10 models generated with different seeds and perform majority voting for predicting base-score values. 
We calculate the \emph{accuracy measure} for all base-scores values, which is the ratio of correct predictions out of the total predictions made. 
The results of training machine learning models are shown in \cref{tab:results-original-features}. 
The accuracy metric listed is on the testing dataset and the best accuracy scores for every base-score value is listed in bold-face text. 
\vspace{1em}

\begin{table}[h]
    \centering
    \caption{Prediction results on test dataset -- with original features (\cref{sec:feature-extraction}) only}
    \label{tab:results-original-features}
    \begin{tabular}{|c|r|r|}
        \hline
        & \textbf{Random Forest} & \textbf{Naive Bayes}                                  \\ \hline
        AV &   \textbf{0.59}          &    0.27           \\
        AC &   0.55          &    \textbf{0.59}           \\
        PR &   0.91          &    \textbf{0.95}           \\
        UI &     \textbf{0.73}        &     0.45          \\
        S  &    \textbf{0.91}         &     0.91          \\
        C  &    \textbf{0.64}         &     0.45          \\
        I  &      \textbf{0.55}       &       0.27        \\
        A  &    \textbf{0.82}         &     0.55         \\ \hline
    \end{tabular}
\end{table}

\begin{table}[ht]
    \centering
    \caption{Summary of feedback received by experts}
    \label{tab:expert-feedback}
    \begin{tabularx}{\linewidth}{|c|X|X|X|} \hline
        \thead{Expert ID} & \thead{Programs \\ analyzed} & \thead{Functions \\ expanded} & \thead{Comments \\ left} \\ \hline
        1 & 1 & 1 & 1 \\
        2 & 1 & 4 & 1 \\
        3 & 1 & 1 & 1 \\
        4 & 1 & 1 & 1 \\
        5 & 1 & 2 & 2 \\
        6 & 1 & 8 & 4 \\
        7 & 3 & 6 & 3 \\ \hline
        \textbf{Unique} & \textbf{5} & \textbf{20} & \textbf{13} \\ \hline
    \end{tabularx}
\end{table}
We can see from these results that, except for \emph{attack complexity} and \emph{privileges required}, random-forest classifier performs the best in terms of test accuracy scores. Even though some of the accuracy scores, such as for \emph{attack vector}, \emph{confidentiality impact} and \emph{integrity impact} seem to be low ($0.59$, $0.64$ and $0.55$ respectively), we don't consider them too bad because these base-score values may be in one of $4$, $3$ and $3$ classes respectively. 

\subsection{Feedback from Experts}\label{sec:eval-feedback}
The best models obtained from the learning phase are used to predict CVSS3 base-scores for the vulnerabilities in the set ($X-G$) (previously unreported vulnerabilities) and the final CVSS3 scores for them calculated using the equations presented in \cite{cvss3}. 
For getting feedback, we contacted 
\begin{enumerate*}
    \item developer mailing-lists of the programs analyzed, 
    \item students of ``Security Engineering'' lecture course, who had sufficient background in secure software development principles and symbolic execution, and
    \item two members of technical staff at our organization, one of whom has a doctoral degree in a security-related field. 
\end{enumerate*}
The call-graphs, respective CVSS3 base-scores, and final scores are, then, shown to the experts using the interface described in \cref{sec:presentation-feedback-gathering}. 

In \cref{tab:expert-feedback}, we have summarized the feedback received from the experts that we contacted. The second column in \cref{tab:expert-feedback} shows the number of programs that an expert analyzed. The third column shows the number of \emph{unique} functions in the programs for which an expert clicked to either, expand to view the source code or, only view the assigned CVSS3 base-scores. The last column in \cref{tab:expert-feedback} shows the number of comments or feedback items left by the respective expert during the entire exercise. The last row of this table lists the number of \emph{unique} analyzed programs, functions and feedback items received by us. 
In \cref{lst:feedback}, we have presented verbatim some of these feedback items. We have omitted some comments from this paper because they were either, identical to the feedback items shown, or were unrelated to bug triage, e.g.\ ``you must examine the latest version of this program because some vulnerabilities were fixed later.''

\lstset{basicstyle=\scriptsize, breaklines=true, numbers=none, showstringspaces=false}
\begin{lstlisting}[language={},caption=Some feedback from experts, label=lst:feedback,emph={Functions,selected, Comment,Program},emphstyle={\bfseries\itshape}]
Program: Jasper 2.0.10
Functions selected: jpc_dec_decodepkt
Comment: 
Without pinpointing the vulnerable instruction, the 
function is very hard to analyze manually. Looking 
at the size of this function (250 lines), it might be a 
good idea to keep the score high, because it's likely 
to be reused somewhere. I also think the signature 
of the function suggests the incoming parameters are 
very varied and, hence, might be prone to being unsanitized

Functions selected: jpc_dec_decodepkt, main, jpc_dec_lookahead
Comment:
The tool looks great, but it would be really useful if 
for each function you would also indicate the number 
of the LOC where the buffer overflow vulnerability 
occurs. Otherwise, for large functions, it is difficult 
to pinpoint the vulnerability manually. Of course, it is 
also easier to analyze the code of commented functions 
in comparison to functions without any comments.

Program: Rzip 2.1
Functions selected: read_buf, write_u16, BZ2_bzBuffToBuffCompress, write_buf
Comment:
All OK.

Functions selected: read_u8
Comment:
All is OK but for this file I'm not sure the result of confidentiality

Functions selected: read_stream, write_stream, write_u32
Comment:
I think the file is used locally.

Program: Libass 0.13.3
Functions selected: ass_pre_blur1_vert_c
Comment: 
To me this function does not seem to be exploitable via the network. 

Program: ImageMagick 6.0.4-8
Functions selected: ReadRLEImage
Comment: 
Here it looks to me like those code will be exploitable via the network as imagemagic is often used to parse network data
\end{lstlisting} 

\subsubsection*{Synthesizing Feedback}\label{sec:synthesizing-feedback}
As a qualitative measure of \emph{effectiveness}, we wanted to know if our framework successfully helped the experts assign severity to vulnerabilities.
We could distill the following main points from the feedback received from experts who used our prediction framework
\begin{enumerate}
    \item Most experts found the tool to be useful.
    \item Triage process is significantly affected by the size of the source-code being analyzed.  
    \item In the absence of relevant comments, the perceived severity of functions is affected by how ``complex'' it is.
    \item The perceived severity of a vulnerable function, somehow, depends on what parameters are passed to it.
    \item Experts would prefer pinpointing of the vulnerable instructions, rather than only the affected function. 
    \item At least one expert was suspicious of the confidentiality impact score assigned to a function.  
\end{enumerate}

\subsection{Machine Learning from Improved Features}\label{sec:eval-improved-features}
Based on the feedback received from experts, we add more features, expecting to increase the accuracy scores for all base-score values. The added features, as listed in \cref{sec:feature-addition}, are -- number of LLVM instructions in the function, number of basic-blocks in the functions and the number of function parameters that are of pointer type. Using these three additional features, we, now, train the same machine learning models as in \cref{sec:eval-feature-extraction-and-machine-learning}, i.e. 4-fold cross-validated random-forest classifier and naive Bayes classifier. The accuracy scores on the test set after adding these three features to the existing features (total \emph{ten} features) are shown in \cref{tab:results-added-features}. 
\begin{table}[h]
    \centering
    \caption{Prediction results on test dataset -- with original and added features (\cref{sec:feature-addition})}
    \label{tab:results-added-features}
    \begin{tabular}{|c|r|r|}
        \hline
        & \textbf{Random Forest} & \textbf{Naive Bayes}                                  \\ \hline
        AV &  \textbf{0.64}           &     0.50          \\
        AC &  \textbf{0.82}           &     0.55          \\
        PR &  \textbf{1.00}           &     0.95          \\
        UI &   \textbf{0.95}          &      0.95         \\
        S  &    \textbf{1.00}         &      0.95        \\
        C  &    \textbf{0.91}         &      0.91         \\
        I  &      \textbf{0.73}       &        0.50       \\
        A  &    \textbf{0.91}         &      0.82        \\ \hline
    \end{tabular}
\end{table}
The results show that, after including three more features the random-forest classifier and naive Bayes classifier were, both, able to more accurately predict \emph{most of} the CVSS3 base-score values of the ground truth. The best scores, as we can see from \cref{tab:results-added-features}, were all found with random-forest classifier. Especially notable is the results that, with new features based on the feedback received from security experts and developers, \emph{privileges-required} (\texttt{pr}) and \emph{scope change} (\texttt{sc}) could be predicted with $100\%$ accuracy in the test dataset. \emph{User-interface} (\texttt{ui}) could also be predicted with an almost-perfect accuracy. 

\subsection{Interpretation of the Results}\label{sec:eval-interpretation}
Based on the various experiments conducted by us, we will now take a big-picture view of the results obtained. 

\subsubsection{Effectiveness of Prediction}
From \cref{tab:results-original-features,tab:results-added-features}, we can see that, while some CVSS3 base-score values could be predicted by our framework with high accuracy, there are others for which the framework does not perform reasonably well. We want to stress in this work that we don't claim that \emph{all features} of extracted functions may correlate with \emph{all base-score values}. 

It is, perhaps, not surprising that the accuracy in predicting \emph{attack vector} and \emph{integrity impact} was not as high as, say, \emph{availability impact}. This is because attack vector of a reported vulnerability describes the means through which a system may be attacked by an outsider. This base-score value may only be predicted based on more detailed information about the deployment, and usage, if at all. Similarly, integrity impact describes the effect on the overall integrity of the data managed, manipulated or protected by the system and it is difficult to predict it based only on features of a function. However, intuitively, the high accuracy in predicting base-score values of \emph{attack complexity} or \emph{availability impact} can be explained, respectively, by taking into account complexity of a function (\cref{sec:feature-addition}) \cite{el2001prediction} and the effect that a buffer-overflow vulnerability would have, if exploited, i.e.\ program crash. This is clearly reflected in our results. 

Therefore, we can claim by looking at our results that our chosen features can be used to assign correctly \emph{most} of the base-score values with a high accuracy for previously reported bugs. However, for other base-score values, where the accuracy of prediction is not as high, we should use other sources, such as function or requirements specifications, or even manual intervention for increasing effectiveness assessment. 

\subsubsection{Effectiveness of Overall Framework}
The comments, as listed in \cref{lst:feedback}, indicate that most experts who used our tool were satisfied by the format of the tool and agreed with the predicted values for base-scores of previously unseen vulnerabilities. Some of the feedback, as seen in \cref{lst:feedback}, was not concerned with the features of the analyzed programs and functions but instead with the presentation of the results. Even though these comments could not be used to enrich the list of features, we used them to improve our interactive tool. For example, we extended it to include and highlight the precise lines of code where a potential buffer-overflow may occur. 

Therefore, by qualitatively analyzing feedback received from the experts, we can claim that such a tool for predicting CVSS3 severities can effectively aid vulnerability assessment. 

\subsubsection{Adaptability of Features}
Our framework depends on effective discovery of vulnerabilities by a reliable tool. We chose Macke for this, due to the following reasons. Firstly, Macke can find \emph{more} vulnerabilities \cite{ognawala2016macke} in isolated functions, than forward symbolic execution. It also generates fewer false-positives than static-analyzers, such as \emph{Splint}. 
Secondly, Macke outputs ``maximum length of infection'' (\texttt{li}), that we discussed in \cref{sec:feature-extraction}. According to an ad-hoc analysis of features, this is one of the most highly correlated features to most CVSS3 base-score values and, therefore, is crucial in increasing the accuracy of predictions. 

However, an important feature of our tool is that it does not depend on the technique used to discover the vulnerable functions. All the features used for machine learning, except \texttt{li}, can be just as easily extracted using any static analyzer or even a manual code review. A comparison of the effectiveness of severity assessment based on different vulnerability scanners is left as a future work. 

\subsection{Threats to Validity}\label{sec:threats-to-validity}
We will now discuss some threats to validity of our results, especially in the context of generalization to other programs or programming languages. 

\subsubsection{Subjectivity of Assessment}\label{sec:threat-assessment-subjectivity}
The subjectivity of existing CVSS3 scores directly affected our methodology. The assignment of these base-scores may not be objective or reproducible when the same vulnerabilities are presented to two different experts. This would be detrimental because we have assumed that our ground truth always holds. We counter this threat by including 21 different programs in our analysis, which introduces variation in factors influencing the manual process. 

\subsubsection{Subjectivity of Feedback}\label{sec:threat-feedback-subjectivity}
Similar to the ground truth for machine learning, there is also subjectivity in feedback from the experts who participated in the study. Since none of the participants analyzed every program, we needed a way to balance their opinions based only on the programs that they analyzed. The measure that we took to counter this threat is to manually interpret the experts' feedback and use a distilled list of feedback items for feature addition. 

\subsubsection{Randomization}\label{sec:threat-randomization}
The machine learning results (accuracy) may not be reproducible with the same initial training and testing dataset, due to the randomization in the implementation of random-forest classifier and naive Bayes classifier in scikit-learn \cite{scikit-learn}. To counter this threat, we performed 10 runs, with different seeds, of every fold of 4-fold verification and obtain the results of prediction from a majority voting algorithm for every base-score value. 

\subsubsection{Criticism of CVSS}\label{sec:threat-cvss}
Another threat to validity is the scoring system of our choice, CVSS3. As discussed previously by \citeauthor{allodi2014comparing} \cite{allodi2014comparing} and \citeauthor{holm2015expert} \cite{holm2015expert}, CVSS scores are not always indicative of how bug fixing in real-world should be prioritized. However, as authors of both these papers admit, there aren't any competing vulnerability ranking measures that take into account as many factors for ranking severity of vulnerabilities as CVSS does and, hence, we chose it for our experiments. 

\subsubsection{External Threats to Validity}
The last threat to validity is the external threat of variance in the \emph{nature} of the analyzed programs. We have analyzed programs for which we could find reported vulnerabilities with CVSS3 scores in NVD. However, the common characteristics of these programs were that they were all open-source and maintained by members of a large and growing community with a different set of skills and expertise that we didn't account for. We claim here that the results obtained by us generalize only over the set of programs analyzed by us, but may not hold for other industrial software, embedded systems or proprietary real-world programs. 
\section{Related Work}\label{sec:related-work}
\subsection{Automatic bug assessment}\vspace{-0.5em}
By observing correctly that most past works in severity prediction \cite{chaturvedi2012determining,yang2014towards,wen2015novel} have been concerned with text-mining from bug repositories, \citeauthor{bettenburg2008makes} \cite{bettenburg2008makes} decided to investigate features that make a bug report \emph{good}. They concluded in this paper that reports containing steps to reproduce bugs and stack traces are considered to be most useful. However, none of the related works analyzed by us make use of the stack traces (that can be automatically generated by executing the exploit) to predict severity. 
\citeauthor{lamkanfi2011comparing} \cite{lamkanfi2011comparing} compared text-mining algorithms and concluded that a naive Bayes classifier performs the best in terms of ROC measure when applied on the textual content of bugs reported in Eclipse and GNOME open-source projects. However, the vulnerabilities could be classified in one of two classes only, viz. \emph{severe} or \emph{not severe}. Research by \citeauthor{menzies2008automated} \cite{menzies2008automated}, and \citeauthor{chaturvedi2012determining} \cite{chaturvedi2012determining} applied similar text-mining approaches on NASA's Project and Issue Tracking System (PITS), with promising results, where the severity scale ranges from 1 to 5, with 1 being most severe.
Other severity prediction frameworks, such as \cite{yang2014towards,wen2015novel,tian2012information,shen2011efindbugs}, have also focussed on the textual content of the bug reports. 
There have been only a few, to our knowledge, works that do severity prediction using source-code or any other intermediate representation (e.g.\ control-flow-graph and call-graph) of the system-under-test. \citeauthor{palomba2016smells} \cite{palomba2016smells} presented a method for improving bug prediction in software components using code smells. \citeauthor{el2001prediction} \cite{el2001prediction} presented a technique for early detection of components in the software that may be faulty, based on metrics derived from the object-oriented design. However, a discussion of the severity of predicted faults is not included in \cite{el2001prediction}. Some past works by \citeauthor{nagappan2006mining} \cite{nagappan2006mining}, \citeauthor{nagappan2005static} \cite{nagappan2005static}, and \citeauthor{el2001prediction} \cite{el2001prediction} have inspired the choice of function and call-graph features that we have used in our study to learn the severity of vulnerabilities in them.

\subsection{CVSS scale}
Papers such as \cite{houmb2010quantifying}, by \citeauthor{houmb2010quantifying}, use CVSS to predict the frequency and impact of failures from a risk management perspective. \citeauthor{dobrovoljc2017predicting} \cite{dobrovoljc2017predicting} suggest an improvement over CVSS that explicitly includes the features of perceived attackers who may be able to exploit a vulnerability. Similarly, \citeauthor{wang2011improved} \cite{wang2011improved}, and \citeauthor{liu2011vrss} \cite{liu2011vrss} also suggest new scales by pointing our certain deficiencies in CVSS that restrict its usage in their contexts. \citeauthor{allodi2014comparing} \cite{allodi2014comparing}, and \citeauthor{holm2015expert} \cite{holm2015expert} present criticisms of CVSS's ability to quantify the severity of discovered vulnerabilities by pointing out that vulnerabilities that are reported with sample exploits can, generally, be a much better indicator of their severities.  

\subsection{Contribution}
By analyzing the related works above our work tries to fill the gaps in research in the following ways --
\begin{enumerate}
    \item Our work treats the system-under-test as a set of interacting components and predicts the severity of vulnerabilities based on heuristics of the affected functions and their interactions with other functions.  
    \item To the best of our knowledge, our framework is the first to correlate call-graph features and compositional analysis results to any severity measures of reported vulnerabilities, CVSS or otherwise.  
    \item Our work also comments on whether certain CVSS base-scores are suitable to be predicted by only the syntactic properties of a program.
\end{enumerate}
\section{Conclusion}\label{sec:conclusion}
In this paper, we described a framework for automatically predicting the severity of vulnerable functions reported by a compositional symbolic execution tool. Using a systematic procedure, we collected data from NVD about vulnerabilities reported in the past with CVSS3 severity scores for C programs. For the collected vulnerabilities, we compiled and analyzed the same programs with Macke, a compositional analysis tool based on symbolic execution. From the results of programs' analyses, we extracted some features for training machine models to predict CVSS3 base-score values. The results from the best model were, then, presented to various experts in the field of secure software development to obtain their feedback on the tool and predicted base-score values. Based on the feedback received, we updated the list of features to be extracted from the programs and re-trained the machine learning models. Our evaluation showed that the accuracy of predictions with the updated list of features had an improvement for all CVSS3 base-score values. Using this empirical result and qualitative feedback from the community of experts, we have shown that our predictive framework can effectively help developers assess the severity of vulnerabilities reported by Macke. 

\bibliographystyle{ACM-Reference-Format}
{\footnotesize
    \bibliography{bibliography,mendeley-symbolic-execution-tag,mendeley-severity-assessment-tag,mendeley-concolic-execution-tag,mendeley-compositional-tag,self-pubs,macke-ase-2016,munch-sac-2018}
}

\end{document}